\documentclass[prl,nofootinbib,preprintnumbers]{revtex4}
\usepackage{graphicx,psfrag,epsfig,epsf,colordvi,slashed}

\begin{document}

\preprint{MAN/HEP/2015/19}
\preprint{December 2015}

\title{{\Large Diphoton Signatures from Heavy Axion Decays\\[0.5mm]
  at the CERN Large Hadron Collider}\\[3mm] }

\author {\large Apostolos Pilaftsis\\}

\affiliation{\vspace{0.2cm} Consortium for Fundamental Physics, School
  of Physics and Astronomy, University of Manchester, Manchester, M13
  9PL, United Kingdom.}

\begin{abstract}
\noindent
Recently, the  LHC collaborations,  ATLAS and  CMS, have  announced an
excess  in  the diphoton  channel  with  local significance  of  about
$3\,\sigma$ around  an invariant mass distribution  of $\sim 750$~GeV,
after  analyzing  new data  collected  at  centre-of-mass energies  of
$\sqrt{s}   =  13~{\rm   TeV}$.   We   present  a   possible  physical
interpretation of such a signature,  within the framework of a minimal
UV-complete model with a massive  singlet pseudo-scalar state $a$ that
couples  to a  new TeV-scale  coloured vector-like  fermion~$F$, whose
hypercharge quantum  number is a non-zero  integer.  The pseudo-scalar
state $a$  might be a  heavy pseudo-Goldstone  boson, such as  a heavy
axion, which  decays into two photons  and whose mass lies  around the
excess region.   The mass  of the  CP-odd state  $a$ and  its coupling
to~$F$ may  be due  to non-perturbative effects,  which can  break the
original Goldstone shift symmetry dynamically.  The possible role that
the heavy axion  $a$ can play in the radiative  generation of a seesaw
Majorana scale and in the solution  to the so-called strong CP problem
is briefly discussed.

\end{abstract}

\maketitle

Recently, the ATLAS and CMS  collaborations have analyzed Run~2 LHC data
gathered at centre-of-mass  energies  of  $\sqrt{s}  =  13~{\rm  TeV}$.   They
reported an  excess in the  diphoton channel around an  invariant mass
distribution of $\sim 750$~GeV, with local significance of $3.6\,\sigma$
and  $2.6\,\sigma$ confidence  level (CL),  respectively~\cite{ATLAS-CMS}.  Interestingly
enough, Run~2 data do not show up any significant excess in  other diboson
channels,  such as  $ZZ$, $W^+W^-$  and $Z\gamma$,  whilst the  Run--1
bumps  seen   around  the   2~TeV  region   have  now   become  almost
statistically insignificant.

In this  short note,  we offer  a possible  interpretation of  such an
excess  in the  diphoton channel,  within the  framework of  a minimal
UV-complete model with a massive  singlet pseudo-scalar state $a$ that
couples  to   new  coloured  vector-like   fermions~$F_{L,R}$.   These
vector-like     fermions     are     very    heavy     with     masses
$m_F \stackrel{>}{{}_\sim}  1.5$~TeV, so as to  have escaped detection
so far  at the  LHC.  They  are charged under  the SU(3)$_C$  group of
Quantum  Chromodynamics~(QCD), but  they are  singlets under  the weak
SU(2)$_L$ group of the Standard Model (SM).  They must also have
non-zero integer  hypercharges, e.g.~$Y_{F_L} =  Y_{F_R} = 1,\, 2,\dots$, which
forbid  them to  have Yukawa  interactions with the SM quarks.   On the
other  hand,  the  pseudo-scalar  state   $a$  may  well  be  a  heavy
pseudo-Goldstone boson, such  as a heavy axion, which  decays into two
photons  with   a  mass  that   lies  around  the  excess   region  of
$\sim 750$~GeV.  Both  the mass $M_a$ of the state~$a$  and its CP-odd
coupling     to     the     Dirac     vector-like     fermion     $F$,
$h_F  \bar{F}i\gamma_5  F$,   could  originate  from  non-perturbative
effects that break the original axion shift symmetry.

The minimal UV-complete  model that we will be  considering here is related
to  the one  that discussed  earlier in  Ref.~\cite{Pilaftsis:2012hq}.
The relevant non-SM part of the  Lagrangian of interest to us is given
by
\begin{equation}
  \label{eq:LU1}
{\cal L}_{\rm } \ =\ \bar{F}\, \Big(i\!\not\!\! D - m_F\Big) F\: 
+\: \frac{1}{2}\,(\partial_\mu a)(\partial^\mu a)\: -\: \frac{1}{2}\,
M^2_a a^2\: -\: h_F\, a\, \bar{F} i\gamma_5 F\; .
\end{equation}
In                              the                             above,
$D_\mu = \partial_\mu + ig_s T^a G^a_\mu  + i g' (Y_F/2) B_\mu$ is the
covariant derivative acting on the  exotic coloured Dirac fermion $F$,
where  $G^a_\mu$ and  $B_\mu$  are the  SU(3)$_C$  and U(1)$_Y$  gauge
bosons,  respectively,  and  $T^a$   (with  $a=1,2,\dots,8$)  are  the
generators   of    the   SU(3)$_C$    gauge   group.     Notice   that
Lagrangian~(\ref{eq:LU1}) is  invariant under the  CP transformations:
$a(t,{\bf     x})     \to    -     a     (t,     -{\bf    x})$     and
$\bar{F}(t,{\bf x} ) i\gamma_5 F(t,{\bf x}) \to - \bar{F}(t, -{\bf x})
i\gamma_5                F(t,               -{\bf                x})$.
In  the   absence  of   the  fermion  mass   term  $m_F   \bar{F}  F$,
Lagrangian~(\ref{eq:LU1})  is  also   invariant  under  the  chirality
discrete       transformations:       $a       \to       -a$       and
$F_{R\,(L)} \to  +(-)\, F_{R\,(L)}$.   Given that  $m_F \neq  0$, this
latter  symmetry is  broken softly  by the  dimension-3 mass  operator
$m_F \bar{F} F$.  Finally, it is  important to remark that the squared
mass    $M^2_a$     and    the     Yukawa    couplings     $h_F$    in
Lagrangian~(\ref{eq:LU1})   break  explicitly   the  Goldstone   shift
symmetry:  $a \to  a +  c$, where  $c$ is  an arbitrary  constant. The
possible origin  of such a  breaking could be due  to non-perturbative
effects related to some unspecified strong dynamics.

\begin{figure}
\includegraphics[scale=0.8]{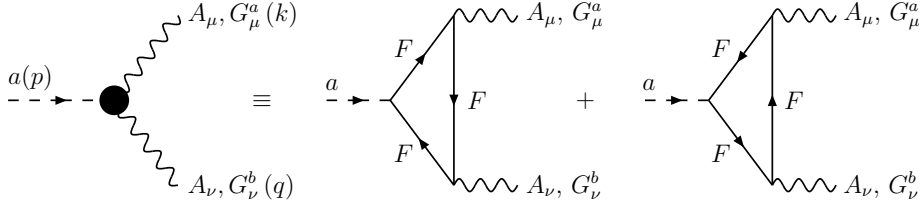} 
\caption{\it  The operators  $a F_{\mu\nu}\widetilde{F}^{\mu\nu}$  and
  $a G^a_{\mu\nu}\widetilde{G}^a_{\mu\nu}$,  as induced by  the chiral
  global  anomaly  of  the  heavy fermion  $F$,  with  the  convention
  $p + k + q = 0$.}\label{fig:chiral}
\end{figure}

In the above  minimal extension of the SM, the  pseudoscalar field $a$
couples to the electromagnetic (em) field~$A_\mu$ and the gluon fields
$G^a_\mu$,       via       the       five-dimensional       operators:
$a\,          F^{\mu\nu}          \widetilde{F}_{\mu\nu}$          and
$a\, G^{a\,\mu\nu}  \widetilde{G}^a_{\mu\nu}$, where  $F^{\mu\nu}$ and
$G^{a,\mu\nu}$ are  the U(1)$_{\rm  em}$ and SU(3)$_C$  field strength
tensors,                       respectively,                       and
$\widetilde{F}_{\mu\nu}               \equiv               \frac{1}{2}
\varepsilon_{\mu\nu\lambda\rho}\,                     F^{\lambda\rho}$
and
$\widetilde{G}^a_{\mu\nu}              \equiv              \frac{1}{2}
\varepsilon_{\mu\nu\lambda\rho}\,                 G^{a,\,\lambda\rho}$
are their corresponding dual tensors.   These operators are induced by
the  chiral global  anomalies of  the heavy  fermion $F$,  through the
triangle graphs  shown in Fig.~\ref{fig:chiral}.  With  the convention
that all  momenta are  incoming, i.e.~$p +  k + q  = 0$,  the one-loop
$a(p)$-$A_\mu(k)$-$A_\nu(q)$                                  coupling
reads~\cite{Steinberger:1949wx,Adler:1969gk,Bell:1969ts}:
\begin{equation}
  \label{eq:aAA}
i\Gamma^{aAA}_{\mu\nu} (p,k,q)\ =\ i\, Q^2_F\, \frac{N_C\,\alpha_{\rm em}}{\pi}\; 
\frac{h_F}{m_F}\; F_P\bigg(\frac{p^2}{4m^2_F}\bigg)\;
\varepsilon_{\mu\nu\lambda\rho}\, k^\lambda q^\rho\; ,
\end{equation}
where $Q_F = Y_{F}/2$ is the electric charge of the heavy fermion~$F$,
$N_C    =     3$    is     its    colour    degrees     of    freedom,
$\alpha_{\rm em}  = e^2/(4\pi)$ is the  electromagnetic fine structure
constant,   and   $\varepsilon_{\mu\nu\lambda\rho}$   is   the   usual
anti-symmetric Levi--Civita  tensor (with the convention: $\varepsilon^{0123}  = +1$).
Moreover, the  loop function  $F_P (\tau )$  was calculated  long time
ago~\cite{Steinberger:1949wx} and found to be:
\begin{equation}
  \label{eq:FPtau}
F_P (\tau ) \ =\  
\left\{ \begin{array}{lr}
\frac{\displaystyle 1}{\displaystyle \tau}\; \arcsin^2 \sqrt{\tau}\; ;
& |\tau| \leq 1\;,\\[2mm] 
-\frac{\displaystyle 1}{\displaystyle 4 \tau}\; \bigg[\, 
\ln\bigg(\frac{\displaystyle \sqrt{\tau} +
    \sqrt{\tau -1}}{\displaystyle \sqrt{\tau} -\sqrt{\tau -1}}\bigg) -
  i\pi\,\bigg]^2\; ; & |\tau| \geq 1\; . \end{array} \right.\qquad
\end{equation}
Note      that     for~$|\tau      |     \ll      1$,     we      have
$F_P(\tau  )  =  1  +  \tau   /3  +  {\cal  O}(\tau^2)$,  whereas  for
$|\tau| \gg 1$,  $F_P (\tau) \to - \ln^2|\tau|/(4\tau)$  which goes to
zero asymptotically as~$\tau \to \infty$.

By  analogy,  the  SU(3)$_C$  global anomaly  generates  an  effective
interaction of  the heavy axion $a$  to gluons $G^a_\mu$, as  shown in
Fig.~\ref{fig:chiral}. The  effective $a(p)$-$G^a_\mu(k)$-$G^b_\nu(q)$
coupling is given by
\begin{equation}
  \label{eq:aGG}
i\Gamma^{aG^aG^b}_{\mu\nu} (p,k,q)\ =\ i\, \delta^{ab}\, \frac{\alpha_s}{2\pi}\; 
\frac{h_F}{m_F}\; F_P\bigg(\frac{p^2}{4m^2_F}\bigg)\;
\varepsilon_{\mu\nu\lambda\rho}\, k^\lambda q^\rho\; ,
\end{equation}
where  $\alpha_s  =  g^2_s/(4\pi  )$  is  the  strong  fine  structure
constant. 

With  the  aid  of  the effective  couplings  given  in~(\ref{eq:aAA})
and~(\ref{eq:aGG}),  it  is  straightforward to  calculate  the  decay
widths of the heavy axion $a$ into photons ($\gamma$) and gluons ($g$):
\begin{eqnarray}
  \label{eq:Widthgamma}
\Gamma (a \to \gamma\gamma) & = &  \frac{N^2_C \alpha^2_{\rm
                                  em}}{64\,\pi^3}\, Q^4_F\, h^2_F\,
                                  \frac{M^3_a}{m^2_F}\: \big| F_P (\tau_a )\big|^2\;,\\
  \label{eq:Widthgluon}
\Gamma (a \to gg) & =&  
\frac{\alpha^2_s}{32\,\pi^3}\, h^2_F\,
                                  \frac{M^3_a}{m^2_F}\: \big| F_P
                       (\tau_a )\big|^2 K^g_a\;,
\end{eqnarray}
where $\tau_a \equiv M^2_a/(4m^2_F) $ and $K^g_a \approx 1.6$ is a QCD
loop  enhancement   factor  which  includes  the   leading  order  QCD
corrections~\cite{Spira:1995rr}. In addition, the other diboson decay
channels, such as $a \to ZZ$, $Z\gamma$ and $W^+W^-$, may be reliably 
estimated to leading order in $M^2_Z/M^2_a$~\cite{Brustein:1999it} as
follows:
\begin{equation}
   \label{eq:Withdiboson}
\frac{\Gamma (a \to ZZ)}{\Gamma (a \to \gamma\gamma )}\ \approx\ 
\frac{\sin^4\theta_w}{\cos^4\theta_w}\ \approx\ 0.082\ ,\qquad 
\frac{\Gamma (a \to Z\gamma)}{\Gamma (a \to \gamma\gamma )}\ \approx\
\frac{2\,\sin^2\theta_w}{\cos^2\theta_w}\ \approx\  0.57\ , 
\end{equation}
whilst  the decay  width  $a \to  W^+ W^-$  is  negligible, since  the
corresponding  $aW^+W^-$  effective  coupling   is  generated  at  the
two-loop  level, e.g.~from  the one-loop  induced $a\gamma\gamma$  
coupling.  To  satisfy the  LHC constraints on  the masses  of exotic
coloured fermions, we  may assume that the vector-like  fermion $F$ is
heavier  than $a$,  e.g.~$m_F\stackrel{>}{{}_\sim} 1.5$~TeV,  in which
case~$\tau_a \ll  1$.  Hence,  the loop function  $F_P (\tau_a  )$ may
well be approximated as $F_P (\tau_a ) \approx 1$.

If we now take the ratio $R$ of the photonic versus the gluonic decay width
given in~(\ref{eq:Widthgamma}) and (\ref{eq:Widthgluon}), we readily find that
\begin{equation}
  \label{eq:R}
R\ \equiv\  \frac{ \Gamma (a \to \gamma\gamma)}{\Gamma (a \to gg) }\
=\ \frac{N^2_C \,\alpha_{\rm em}^2 Q^4_F}{2\,\alpha^2_s\, K^g_a}\ .
\end{equation}
Observe that the ratio $R$ is independent of the Yukawa coupling $h_F$ and,
for $Q_F \ge 2$, we obtain $R > 1$ and the decay $a \to \gamma\gamma$
can easily become the dominant mode.

The  production   cross  section  of  heavy   axions  via  gluon-gluon
fusion~\cite{Jaeckel:2012yz} may be calculated as follows:
\begin{equation}
  \label{eq:sigma}
\sigma (pp \to a \to \gamma\gamma ) \ \approx\ \sigma (pp
\to a )\: B(a \to \gamma\gamma )\; ,  
\end{equation}
where $B(a \to \gamma\gamma ) \approx R/(1 + 1.57 R)$ is the branching
fraction for  the decay  channel $a\to  \gamma\gamma$, with  $R$ given
in~(\ref{eq:R}).  For centre-of-mass energies  of $\sqrt{s} = 13~{\rm TeV}$,
we may naively estimate the cross section $\sigma (pp \to a)$ as
\begin{equation}
\sigma (pp \to a) \ \sim\  \sigma_{\rm SM} (pp \to H)\times h^2_F\, 
\frac{m^2_t}{m^2_F}\, \frac{M^2_a}{M^2_H}\ ,
\end{equation}
where $\sigma_{\rm SM} (pp \to H)  \approx 40~{\rm pb}$ is a reference
production cross  section of  the SM Higgs  boson $H$  via gluon-gluon
fusion, with  $M_H \approx 125$~GeV~\cite{Agashe:2014kda}.  Hence, for
$M_a = 750$~GeV (or $M_a/M_H = 6$), $m_F/m_t = 10$ and $h_F = 0.1$, we
find that
\begin{equation}
\sigma (pp \to a \to \gamma\gamma ) \ \sim\ 15~{\rm fb} \times B( a
\to \gamma\gamma)\; .
\end{equation}
For $B  ( a \to  \gamma\gamma ) \sim  1$ and an  integrated luminosity
${\cal L} =  3~{\rm fb}^{-1}$ at $\sqrt{s} = 13$~TeV,  we obtain about
45  signal  events, which is compatible with the  diphoton-excess events
reported in~\cite{ATLAS-CMS}.

As discussed  in detail in~\cite{Pilaftsis:2012hq},  axion-like fields
could act  as mediators  to generate TeV-scale  gauge-invariant masses,
such  as  $m_F\bar{F}  F$,  for vector-like  fermions  through  global
anomalies at  the three-loop level.  In  particular, a gauge-invariant
Majorana   mass   term    $m_M   (\overline{\nu}_R)^C\nu_R$   can   be
generated~\cite{Mavromatos:2012cc},  if heavy  axion fields  couple to
Kalb--Ramond  axions~\cite{Kalb:1974yc,Bowick:1988xh}  that  occur  in
torsionful  theories  of  Quantum   Gravity.   Light  axions  play  an
important  role in  solving the  strong CP  problem via  the so-called
Peccei--Quinn       mechanism~\cite{Peccei:1977hh,      Peccei:1977ur,
  Weinberg:1977ma, Wilczek:1977pj}.  Thus, the  possible presence of a
heavy         axion,         or         a         multitude         of
axions~\cite{Arvanitaki:2009fg,Cicoli:2012sz},   may   give  rise   to
interesting  mixing   phenomena  and   possibly  to  new   effects  in
astrophysical considerations~\cite{Tinyakov:2015cgg}.

In conclusion, we have presented a minimal UV-complete model, based on
the    possible   existence    of    a   heavy    axion   with    mass
$M_a  \approx   750$~GeV,  which  could  offer   a  possible  physical
interpretation of the diphoton excess observed in the Run~2 data.  The
model requires  the presence of  a new TeV-scale  coloured vector-like
fermion~$F$,  which  has a  non-zero  integer  hypercharge. For  large
electric   charges   $Q_F   \ge    2$,   the   photonic   decay   mode
$a\to  \gamma\gamma$  becomes  naturally the  dominant  channel.   The
latter,     along    with     the     branching    fractions     given
in~(\ref{eq:Withdiboson}), provide a unique prediction of our minimal
model that  can be  tested with future  Run~2 data.  Nevertheless, our
model may require an extension to  its field content, as it exhibits a
Landau pole at energy scales $Q\stackrel{<}{{}_\sim} 10^{14}$~GeV, for
$Q_F \ge 2$. Further studies are therefore needed, so as to be able to
fully  assess  the  physical  significance of  the  observed  diphoton
excess, as a firm signature of New Physics at the LHC.

\bigskip

\subsection*{Acknowledgements} 
The  author thanks  Andrew  Pilkington for  discussions  on the  ATLAS
signal   events.    This   work   is  supported   in   part   by   the
Lancaster--Manchester--Sheffield  Consortium for  Fundamental Physics,
under STFC research grant: ST/L000520/1.

\end{document}